\documentclass[conference]{IEEEtran}
\IEEEoverridecommandlockouts

\usepackage{adjustbox}
\usepackage[linesnumbered,algo2e,ruled]{algorithm2e}
\usepackage{multirow}
\usepackage{booktabs,subcaption,dcolumn}
\usepackage{tikz}
\usepackage{tabularx}
\usepackage{stfloats}
\usepackage{subcaption}
\usepackage{algorithmic}
\usepackage{textcomp}

\usepackage{balance}
\usepackage{hyperref}

\makeatletter
\def\url@leostyle{%
  \@ifundefined{selectfont}{\def\UrlFont{\sf}}{\def\UrlFont{\small\ttfamily}}}
\makeatother
\newcommand\res[2]{{$#1$}\tiny{\text{$\pm #2$}}}
\newcommand\bestres[2]{{$\mathbf{#1}$}\tiny{\text{$\mathbf{\pm #2}$}}}

\ifdefined\showchanges
\newcommand{\changed}[1]{\color{blue}{#1}\color{black}\xspace}
\else
\newcommand{\changed}[1]{#1\xspace}
\fi

\usepackage{array}

\usepackage{cite}
\usepackage{amsmath,amssymb,amsfonts}
\usepackage{algorithmic}
\usepackage{graphicx}
\usepackage{textcomp}
\usepackage{xcolor}
\def\BibTeX{{\rm B\kern-.05em{\sc i\kern-.025em b}\kern-.08em
    T\kern-.1667em\lower.7ex\hbox{E}\kern-.125emX}}
\begin{document}

\title{Are We All in a Truman Show? Spotting Instagram Crowdturfing through Self-Training
}

\author{\IEEEauthorblockN{
Pier Paolo Tricomi\IEEEauthorrefmark{1}\IEEEauthorrefmark{3},
Sousan Tarahomi\IEEEauthorrefmark{2},
Christian Cattai\IEEEauthorrefmark{1},
Francesco Martini\IEEEauthorrefmark{1},
Mauro Conti\IEEEauthorrefmark{1}\IEEEauthorrefmark{3}
}

\IEEEauthorblockA{\IEEEauthorrefmark{1}Department of Mathematics, University of Padua, Padua, Italy \\\{ tricomi,conti\}@math.unipd.it \{christian.cattai,francesco.martini.11\}@studenti.unipd.it}
\IEEEauthorblockA{\IEEEauthorrefmark{2}Department of Computer Science, University of Twente, Enschede, Netherlands -- s.tarahomi@utwente.nl}
\IEEEauthorblockA{\IEEEauthorrefmark{3}Chisito S.r.l., Padua, Italy}
}



\maketitle

\begin{abstract}
Influencer Marketing generated \$16 billion in 2022. Usually, the more popular influencers are paid more for their collaborations. Thus, many services were created to boost profiles' popularity metrics through bots or fake accounts. However, real people recently started participating in such boosting activities using their real accounts for monetary rewards, generating ungenuine content that is extremely difficult to detect. To date, no works have attempted to detect this new phenomenon, known as crowdturfing (CT), on Instagram.  

In this work, we propose the first Instagram CT engagement detector. Our algorithm leverages profiles' characteristics through semi-supervised learning to spot accounts involved in CT activities. Compared to the supervised approaches used so far to identify fake accounts, semi-supervised models can exploit huge quantities of unlabeled data to increase performance. We purchased and studied 1293 CT profiles from 11 providers to build our self-training classifier, which reached 95\% F1-score. We tested our model in the wild by detecting and analyzing CT engagement from 20 mega-influencers (i.e., with more than one million followers), and discovered that more than 20\% was artificial. We analyzed the CT profiles and comments, showing that it is difficult to detect these activities based solely on their generated content.

\end{abstract}


\section{Introduction}

\begin{figure}[!h]
    \centering
    \subfloat[\centering Fake Profile]{
        \includegraphics[width=0.4\linewidth,height= 3.2cm]{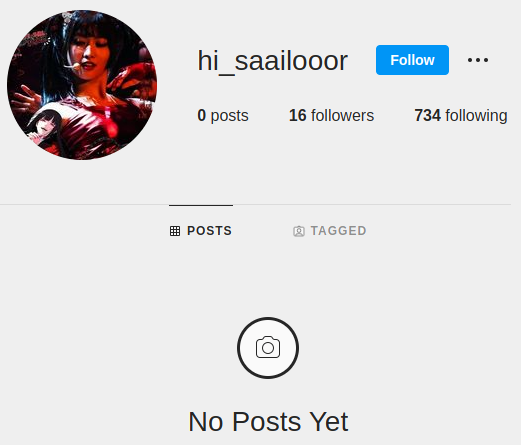}
        \label{fig:fake_ex} }
    \subfloat[\centering Crowdturfing (CT) Profile]{
        \includegraphics[width=0.4\linewidth,height= 3.2cm]{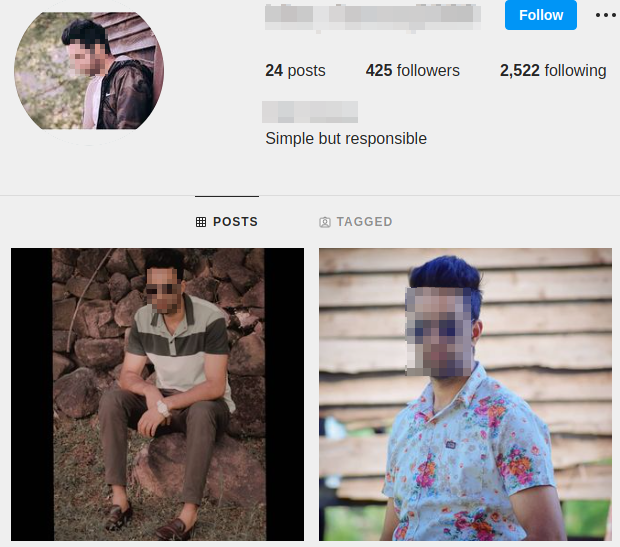}
        \label{fig:ct_ex} 
        }%
        
    \caption{Example of fake vs crowdturfing profiles.}
    \label{fig:fake_CT}%
\end{figure}

Instagram (IG) is the most popular photo-sharing social media, with around 1.5 billion monthly active users~\cite{twitignumber}, and the preferred platform for influencer marketing~\cite{twitiginf}. 
Unfortunately, such a market is often manipulated, making influencers unreliable~\cite{liao2021should}. Indeed, many providers offer services to boost the visibility and fame of a specific account, for example, by increasing the number of followers, likes, and comments. As (social) bots~\cite{cresci2020decade} or fake accounts~\cite{roy2020fake} originally conducted these activities, IG has adopted Machine Learning (ML) algorithms to remove them efficiently. Instead, nowadays, \textit{real people use their accounts to conduct such unauthentic activities behind a monetary reward}. In the literature, this collusive phenomenon is called crowdturfing (CT), a term combining the collaboration of many individuals (crowdsourcing~\cite{abbas2022goal}) with an apparently natural action controlled by agencies (astroturfing~\cite{wang2012serf, elmas2021ephemeral}). 
Figure~\ref{fig:fake_CT} shows Fake and CT profiles. While the fake profile exhibits well-known characteristics (e.g., no posts, no bio, few followers~\cite{roy2020fake}), the CT profile looks legit (indeed, it is a real-person account), and thus more challenging to spot. 
Considering CT engagement is not real, we can label it as fake. Fake engagement damages the authenticity of social media, creating threats such as brand abuse or followers farming~\cite{zarei2020impersonators}. 


To spot fake engagement derived from CT, a reliable strategy is to detect the involved accounts. 
Different approaches have been proposed to distinguish between genuine or fake users, but to the best of our knowledge, none consider CT-involved accounts on IG. Among these approaches, machine learning-based solutions are the most powerful and cost-effective techniques~\cite{orabi2020detection}. While most proposed ML algorithms for account classification leverage supervised learning, there is always the need for an adequately labeled ground truth, which is inherently difficult to obtain for CT activities~\cite{song2015crowdtarget}. %
Instead, Semi-Supervised Learning (SSL) methods could be more appropriate when only a few labeled samples are available or needed. In fact, a large amount of unlabeled data can help improve the classification without impacting the performance~\cite{shi2017semi}. Last, given the intrinsic differences between IG and other social media where CT has been studied (e.g., Twitter), we run our detector in the wild to analyze CT profiles and their fake engagement under several aspects, highlighting the difficulty of detecting such activities merely by looking at generated content.

\textbf{Contribution.} 
Our contribution is summarized as follows:
\begin{itemize}
    \item We are the first, to the best of our knowledge, to propose a CT engagement detector on IG, which furtherly reduces the computational costs of previous fake accounts and bot detectors by leveraging semi-supervised algorithms;
    \item We provide a detailed analysis of CT providers to explore the services they offer and the profiles involved;
    \item We analyze Instagram CT engagement in the wild, mainly related to comments, by running our detector on 1000 posts generated by 20 mega-influencers. 
\item Our (anonymized) data will be released upon request to help researchers study CT activities on IG.
\end{itemize}
\textbf{Structure.} §\ref{sec:related} presents related works. §\ref{sec:providers} examines CT providers, while §\ref{sec:experiments} describes our CT detection mechanism. CT profiles and comments spotted in the wild are analyzed in §\ref{sec:profile_analysis} and §\ref{sec:comment_analysis}, respectively. §\ref{sec:conclusion} concludes the paper. 
\section{Related Works}
\label{sec:related}

We discuss CT detection in social media(§\ref{subsec:ct}), along with fake account detection on IG (§\ref{subsec:fake_and_CT}) and the adoption of semi-supervised algorithms to detect bots and fake profiles (§\ref{subsec:semi-sup}).

\subsection{Crowdturfing in Online Social Media}\label{subsec:ct}
Researchers have examined CrowdTurfing (or \textit{collusive}~\cite{zarei2020impersonators}) social media activities first on Sina Weibo, and then primarily on Twitter, on which misinformation 
or political campaign manipulation 
often occurs.
Wang et al.~\cite{wang2012serf} investigated two popular crowd-sourcing sites in china and tracked down the CT campaigns on Sina Weibo. Then, they discussed the characteristics of CT and genuine accounts and analyzed the CT campaigns. 
Another work on Sina Weibo~\cite{yang2015penny} examined CT accounts engaging in political activities,  claiming their methodology could not find any clear evidence to show the presence of large-scale political CT.
The authors of~\cite{lee2014dark} categorized different types of CT tasks on Fiverr and applied ML algorithms to distinguish these tasks from legitimate ones. 
Song et al.~\cite{song2015crowdtarget} focused on spotting targets of CT tasks, such as pots, pages, and URLs, on Twitter. 
Chetan et al.~\cite{chetan2019corerank} proposed CoReRank,  an  unsupervised  method  for detecting suspicious tweets and collusive retweeters. Dutta et al. developed several mechanisms to detect and characterize collusive users involved in black market services on Twitter~\cite{dutta2019blackmarket,dutta2021decife,dutta2021abome}. 
Eventually, the authors in~\cite{voronin2018crowdturfing} qualitatively investigated the impact of CT activity on content visibility and popularity on IG. They claimed that IG is vulnerable to CT activities and stressed the need for a CT detector. 
To the best of our knowledge, we are the first to implement such a detector for IG, adopting a performing and efficient SSL approach.

As outlined in~\cite{dutta2022blackmarket}, social media have unique characteristics, purposes, and interactions that require tailored CT studies. For instance, researchers have recently moved their interests to YouTube~\cite{dutta2021detecting, dutta2022weakening}, showing that platforms besides Twitter need to be studied. We argue that IG is fundamentally different from Twitter. First, IG has roughly 1.5 Billion monthly active users (three times Twitter ones~\cite{twitignumber}), who spend three times the time spent on Twitter~\cite{twitigtime}, indicating IG's greater influence (2022). Second, they serve very different purposes~\cite{manikonda2016tweeting}: Twitter lets users communicate in an elevator pitch fashion with quick messages, while IG primarily focuses on creating interactive communities through images and videos. Not surprisingly, nearly 80\% of brands use IG influencers for their marketing campaigns, compared to 20\% on Twitter~\cite{twitiginf, hernandez2020towards}. Last, while Twitter APIs\footnote{\url{https://developer.twitter.com/en/docs/twitter-api}} allow collecting a variety of users' data (e.g., profile info, activities, connections), IG APIs\footnote{\url{https://developers.facebook.com/docs/instagram-api/}} release only limited data. Due to these reasons, algorithms developed on Twitter cannot inherently apply to IG, so deploying methods to detect IG CT activities is urgently needed.


\subsection{Instagram Fake Accounts Detection}
\label{subsec:fake_and_CT}
Although no prior works attempted to detect CT activities on IG, several works tried detecting fake profiles~\cite{roy2020fake} or (social) bots\cite{stieglitz2018social, cresci2020decade}, which we can refer to as \textit{classic} fake profiles. 
In~\cite{thejas2019learning}, the authors developed an ML model to detect fake likes on IG, deploying honeypots and botnets to collect the ground truth. They employed ML methods to find the authenticity of likers with features including the number of followers, following, and their relationships. 
To detect fake and automated IG accounts, the authors in~\cite{akyon2019instagram} applied different ML algorithms on posts and media-related features, obtaining 86\% and 94\% accuracy for automated and fake accounts, respectively. 
In~\cite{sheikhi2020efficient}, the authors used bagged decision trees on profile-related features to detect trivial (manually labeled) fake users. 
Zarei et al.~\cite{zarei2020impersonators} applied clustering methods 
to track down impersonators in three different categories based on their profile similarity. 
In~\cite{purba2020classification}, the authors tried to detect three categories of fake accounts: active, inactive, and spammers. They bought fake accounts from Indonesian providers; however, most were simple bots, not linked to CT activities. They reached 92\% accuracy using Random Forest. 
Kim and Hany~\cite{kim2020detecting} proposed a neural network to detect engagement bots by three sets of features, including text, behavior, and graph-based features.
Given the existence of fake accounts and bots detection mechanisms in the literature, we will evaluate such methods on CT profiles, understanding to which extent \textit{classic} fake accounts differ from CT accounts. 

\subsection{Semi-supervised Fake Accounts Detection}
\label{subsec:semi-sup}
SSL approaches can leverage a vast amount of unlabeled data, reducing labeling costs with few to no drops in performances~\cite{shi2017semi}.
Most of these approaches were adopted on social media to detect Sybil Nodes or Bots. 
SybilBelief~\cite{gong2014sybilbelief} is an SSL framework for finding Sybil nodes such as spammers and impersonators. 
SybilTrap~\cite{al2018sybiltrap} uses label propagation random walk as a semi‐supervised transductive‐learning approach to detect malicious users. This approach focuses on both structural and content-based features. 
Dorri et al.~\cite{dorri2018socialbothunter} developed SocialBotHunter as an SSL collective classification technique to detect social bots in Twitter-like platforms. Their approach uses the social behavior and interaction of users. 
Last, SEMIPSM~\cite{alvari2021semi} is an  SSL Laplacian SVM model using manifold regularization to discover users responsible for propagating misinformation on social media.

\section{Crowdturfing Providers Analysis}
\label{sec:providers}
\begin{table*}[!ht]
\scriptsize
\caption{Characteristics of Crowdturfing providers. The table reports information claimed by the provider and retrieved by analyzing 100 profiles bought from each. The last row reports info on real profiles for comparison.}
\label{tab:fake_providers}
\begin{tabular}{ccccccccccc}
\toprule 
\textit{\textbf{Provider}} & \textit{\textbf{Price}} & \textit{\textbf{\begin{tabular}[c]{@{}c@{}}Delivery \\ Time\end{tabular}}} & \textit{\textbf{\begin{tabular}[c]{@{}c@{}}Drop \\ Protection\end{tabular}}} & \textit{\textbf{\begin{tabular}[c]{@{}c@{}}Followers \\ Received\end{tabular}}} & \textit{\textbf{\begin{tabular}[c]{@{}c@{}}Followers \\  1 Month\end{tabular}}} & \textit{\textbf{\begin{tabular}[c]{@{}c@{}}\#Followers\\ Avg (std)\end{tabular}}} & \textit{\textbf{\begin{tabular}[c]{@{}c@{}}\#Following\\ Avg (std)\end{tabular}}} & \textit{\textbf{\begin{tabular}[c]{@{}c@{}}Private\\ Profiles\end{tabular}}} & \textit{\textbf{\begin{tabular}[c]{@{}c@{}}\#Posts\\ Avg (std)\end{tabular}}} & \textit{\textbf{\begin{tabular}[c]{@{}c@{}}URLs in\\ Biography\end{tabular}}} \\
        \midrule
CT-1 & \$5.69 & Instant & yes & 115 & 74 & 409.59 (1110.46) & 812.38 (1331.52) & 0.13\% & 14.83 (57.98) & 0.08\% \\
CT-2 & \$2.39 & 5-10m & no  & 211 & 340 & 44.61 (106.85) & 4679.75 (1452.19) & 0\% & 16.0 (8.06) & 0\% \\
CT-3 & \$2.95 & Instant & yes  & 111 & 85 & 132.17 (327.28) & 3027.08 (1883.18) & 0.05\% & 20.19 (55.99) & 0.09\% \\
CT-4 & \$2 & Instant & no & 100 & 42 & 239.45 (262.64) & 2735.6 (1286.65) & 0.45\% & 111.95 (332.2) & 0.01\% \\
CT-5 & \$3.95 & Gradual & yes & 79 & 61 & 201.43 (214.0) & 3510.77 (2316.12) & 0\% & 16.06 (12.13) & 0.054\% \\
CT-6 & \$2.89 & 24-72h & yes  & 136 & 129 & 36.79 (39.64) & 2398.88 (2191.18) & 0\% & 14.06 (5.69) & 0\% \\
CT-7 & \$2.70 & 1h & yes & 108 & 109 & 39.23 (73.32) & 3966.36 (761.16) & 0\% & 19.74 (20.13) & 0\% \\
CT-8 & \$5.78 & Gradual & no  & 110 & 95 & 57.52 (138.97) & 1818.84 (1353.95) & 0.04\% & 29.75 (41.09) & 0.01\% \\
CT-9 & \$3.95 & 12h & no  & 109 & 99 & 129.54 (759.85) & 2012.93 (1198.17) & 0.06\% & 26.99 (74.94) & 0\% \\
CT-10 & \$5.94 & Gradual & no  & 97 & 94 & 83.38 (174.57) & 2118.31 (1323.78) & 0.03\% & 40.28 (51.5) & 0\% \\
Low quality & \$0.80 & 24-72h  & no & 117 & 96 & 87.26 (276.26) & 3200.67 (3041.89) & 0.04\% & 1.88 (6.15) & 0.02\% \\
\textbf{Real} & - & - &  - & - & - & 359.33 (237.87) & 571.24 (517.53) & 57.92\% & 279.09 (369.67) & 14.44\%
\\\bottomrule
\end{tabular}
\vspace{-1.5em}
\end{table*}

To spot CT activities, such as fake followings or comments, we must study, understand, and collect ``authentic'' CT profiles. Previous studies on fake profiles detection collected fake profiles or bots by manually searching for poorly designed accounts, such as those without a profile pic, with alpha-numeric names, or a very low number of posts and followers~\cite{akyon2019instagram,sheikhi2020efficient}. Other works focused on synthetic data~\cite{thejas2019learning} or bought mostly naive fake accounts from local providers~\cite{purba2020classification}. However, the profiles gathered using these methodologies indubitably introduce bias in the data, and the resulting detectors will identify just simple profiles, very likely driven by a bot master. 
\par
Instead, we are interested in spotting fake activities conducted by real people profiles that are populating and remaining on IG by evading its bot detection mechanisms~\cite{ig_detectors, ig_authenth, ig_bots}. To this aim, we selected 10 well-known crowdturfing providers and bought from each of them 100 fake followers. \changed{All the selected providers \textbf{ensure to deliver real followers (i.e., real people)} who interact with the target profiles by liking and commenting on their posts to boost their engagement rate. These CT profiles are identified as high-quality followers and usually cost more than ``base'' fake profiles (i.e., profiles usually managed by a bot master). To identify reliable providers, we selected services that had at least an average of 3 (out of 5) stars on the famous reviews platforms TrustPilot\footnote{\url{https://www.trustpilot.com/}}. Moreover, many of our CT providers allow people to freely join their platforms to participate in CT activities, confirming the reliability of the service and the presence of human activity behind the fake engagement they generate. }
Table~\ref{tab:fake_providers} describes these providers, along with information about a low-quality provider. We also included information on real profiles we used in our study.\footnote{Here, to simplify comparisons, we excluded celebrities and highly-followed accounts ($>$ 500 followers), which could present inflated statistics.} To limit CT activities on IG, we bought CT followers for profiles we created for the study, which we deleted at the end. We are not reporting the names of the CT providers to avoid the encouragement of such activities. 
\par
The table shows that the price average is pretty low, around \$3 for 100 high-quality followers, but much higher than the \$0.80 for 100 low-quality followers. Followers are usually delivered within a few hours, and most providers offer drop protection, replenishing any lost follower. To assess the providers' reliability, we checked how many followers remained after one month. On average, we lost only 15-20\% of them, and sometimes, we gained more. CT-4, the least expensive provider, lost the most, while we lost only 3 followers from the most expensive CT-10. Compared to real profiles, CT profiles have a noticeable difference in followers and following. This is understandable, given that the more they follow and interact, the more they get paid. However, from CT-1, the second most expensive provider, the follower/following balance is quite close to real profiles. The CT profiles also are quite different from real ones in terms of being private, the number of posts, and the URLs in the biography. Very likely, CT platforms require profiles to be public. People joining these platforms generate a minimum amount of posts to be reliable, except few cases (CT-4, CT-10). The low-quality profiles show a very high imbalance in followers and following, and the average number of posts is close to 0, far below the CT profiles. Among the properties we did not report in the table, some providers allow customers to increase their followers periodically or buy followers from a specific geographical region or language. 

\section{Crowdturfing Profiles Detection}
\label{sec:experiments}
\changed{
Instead of directly detecting CT activities (e.g., a fake comment), we first detect profiles involved in CT activities, and accordingly, we label their interaction as CT. The rationale behind this approach is that CT profiles are mainly \textbf{real accounts} belonging to individuals willing to create fake interactions. Thus, their interactions should resemble genuine ones, in both content and temporal activity~\cite{dutta2022blackmarket}. Similarly, their profile information should appear legitimate, which makes detecting CT profiles considerably different from spotting \textit{classic} fake profiles~\cite{dutta2021decife} (i.e., the focus of previous works). Indeed, the latter usually present simplistic features (e.g., no posts, no followers), or recognizable patterns (e.g., low-variability content)~\cite{roy2020fake}. 
We now present the dataset we collected to classify CT profiles (§\ref{subsec:dataset}), our detection model (§\ref{subsec:model_sel}), and a comparison with previous approaches (§\ref{subsec.baseline}). 
}

\subsection{Dataset and feature selection}
\label{subsec:dataset}
Since there are no IG CT datasets available, we collected our own. Given IG API could not provide our requirements, we performed automated data collection through Selenium\footnote{\url{https://www.selenium.dev/}}. 
For our detector, we use general profile info (e.g., \#followers, \#following, \#posts) instead of behavioral patterns since IG does not provide such information, unlike other social media (e.g., Twitter). Some previous works~\cite{thejas2019learning, purba2020classification} used features that are not publicly available, e.g., the number of likes of a user's posts, limiting their approach only to public profiles. Instead, we focused only on profile features that are publicly available for both public and private profiles.\footnote{Some attributes (e.g., \#videos) were retrieved from the page source code.}

The dataset contains the profile information of 2600 users, including 1293 CT and 1307 authentic accounts. The CT profiles are the ones analyzed in Section~\ref{sec:providers}. We gathered authentic accounts similarly to previous works~\cite{sheikhi2020efficient, akyon2019instagram, purba2020classification}. We included from several countries and fields: general users from our expanded social connections, verified or business accounts, and celebrities. Three authors validated these accounts through extensive manual labeling, adopting a majority voting for the decision, and focusing on attributes such as the Follower/Following imbalance, the number of posts, or the full name. The feature distributions of our real accounts (Table~\ref{tab:fake_providers}) closely align with previous works. 
For the collected accounts, we gathered all the attributes available on the profile page. 
Then, we pre-processed the features by removing those with zero or very low variance. Last, we transformed categorical and non-numeric attributes into numeric or boolean features. The final features are shown in Table~\ref{tab_features}. Since all the data we collected is public, we will make it available upon request (anonymized) to help the research community studying CT. 

\begin{table}[htbp]
\caption{Final set of features of our dataset.}
\vspace{-1em}
\begin{center}
\resizebox{0.9\columnwidth}{!}{
\begin{tabular}{cc}
\toprule
 \textbf{\textit{Numeric Features}}& \textbf{\textit{Boolean Features}}\\
\midrule
 \# followers, \#following & Account is private\\ 
 \# videos, \#posts & Account is verified\\ 
\# char in username, \#digit in username & Account has clips\\ 
\# characters in fullname & Account is business account\\ 
\# characters in biography & Account has external URLs\\ 
\# non-alphabetic char in fullname & Accounts has category name\\ 
\#  hashtags and mentions in biography  & Account has multiple categories\\

 \bottomrule
\end{tabular}
\label{tab_features}
}
\vspace{-1.5em}
\end{center}

\end{table}

\vspace{-0.6em}

\subsection{Our Semi-Supervised Model}
\label{subsec:model_sel}
\begin{figure}[!ht]
    \centering
    \includegraphics[width=0.9\columnwidth]{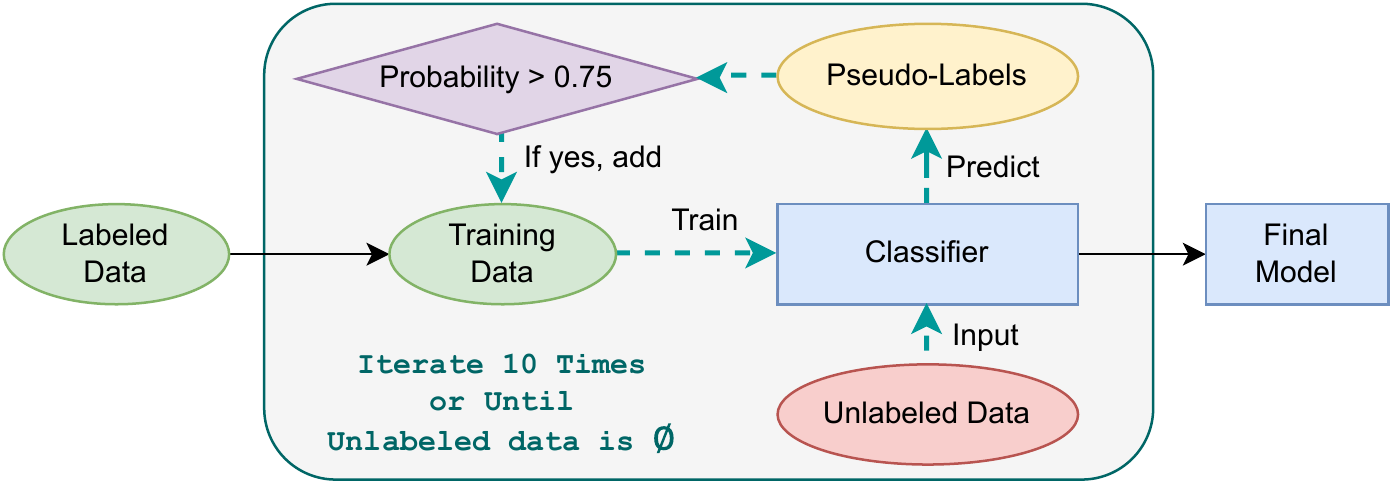}
    \vspace{-0.5em}
    \caption{Schema of Self-Training process. Dashed arrows represent the training cycle.}
    \label{fig:self_tr}
\end{figure}

The next step is to develop a detector to distinguish between real and CT profiles.
In light of previous discussions, labeling CT profiles is challenging~\cite{song2015crowdtarget}. Therefore, instead of adopting supervised methods as in previous works, we use SSL to maximize the use of unlabeled data and improve generalization. 
While most previous SSL approaches on social networks utilized graph-based methods, we consider only profile-related features, making our model less complicated and easier to handle (e.g., for practitioners). 
Our self-training approach is depicted in Figure~\ref{fig:self_tr}. 
\changed{
Initially, we divide the dataset into labeled and unlabeled datasets, discarding all labels from the unlabeled dataset. In the first training cycle (dashed arrows in the figure), training data corresponds to labeled data. We train a classifier with this data and ask it to predict the labels of all the unlabeled samples, generating their pseudo-labels. For each pair \textit{sample:pseudo-label}, we check the prediction probability (i.e., classifier confidence, from 0 to 1) associated with the pseudo-label. If the probability is higher than 0.75, we add the pair \textit{sample:pseudo-label} to the training data; otherwise, the sample remains unlabeled. We repeat the training cycle (train the classifier $\rightarrow$ predict pseudo-labels $\rightarrow$ enlarge the training set) 10 times or until no unlabeled data remains. The final model corresponds to the classifier of the last iteration.
} 


\vspace{-1em}

\changed{
We implemented our models using Scikit-learn.\footnote{\url{https://scikit-learn.org}} 
We randomly split our dataset in a stratified mode to have 80 percent of data for training (2080 labeled samples) and the remaining for testing. We applied 5-StratifiedKFold Cross-Validation on the training set to find the best model hyper-parameters. In each iteration, one fold ($\sim$416 samples) was left out to validate the model, while the remaining $\sim$1664 samples were used to train the classifier in the semi-supervised fashion described above.   
To demonstrate the power of SSL, we tested different (small) labeled training data portions: 1, 3, 5, and 9\%.  As classifiers and hyper-parameters, we tested:
\begin{itemize}
    \item K-Nearest Neighbor (KNN): \textit{n\_neighbors}=[1, 3, 5, 10];
    \item Logistic Regression (LR): \textit{penalty}=[none, l1, l2], \textit{C}=[10, 1, 0.1], \textit{solver}= [lbfgs, liblinear];
    \item Decision Tree (DT): \textit{max\_depth}=[none, 3, 5, 10], \textit{samples\_leaf}=[1, 3, 5, 10];
    \item Random Forest (RF): \textit{max\_depth}=[none, 3, 5, 10], \textit{samples\_leaf}=[1, 3, 5, 10], \textit{n\_estimators}=[10, 100].    
\end{itemize}
The best hyper-parameters for each model were selected through a grid-search approach. 
}
We also trained the classifiers on all the labeled data (i.e., in a supervised mode) for comparison. The results for each classifier using the best hyper-parameters during the cross-validation are reported in Table~\ref{tab_model}.

\begin{table}[htbp]
\caption{Average$\pm$std of classification results of the best models during cross validation. Sup = Supervised.}
\vspace{-1em}
\begin{center}
\resizebox{\columnwidth}{!}{
\begin{tabular}{ccccccc}
\toprule
 \multicolumn{2}{c}{\textbf{\textit{\begin{tabular}[c]{@{}c@{}}Model and \\\% Labels Used \end{tabular}}}}&\textit{\textbf{\begin{tabular}[c]{@{}c@{}}Train\\Accuracy\end{tabular}}}& \textit{\textbf{\begin{tabular}[c]{@{}c@{}}Valid. \\ Accuracy\end{tabular}}}& \textit{\textbf{\begin{tabular}[c]{@{}c@{}}Valid. \\ Precision\end{tabular}}}& \textit{\textbf{\begin{tabular}[c]{@{}c@{}}Valid. \\ Recall\end{tabular}}}& \textit{\textbf{\begin{tabular}[c]{@{}c@{}}Valid. \\ F-Measure\end{tabular}}}\\
\midrule
 &\textit{\textbf{0.01}}& \res{0.79}{0.04}& \res{0.79}{0.03}& \res{0.84}{0.02}& \res{0.79}{0.04}& \res{0.78}{0.04}\\
 &\textit{\textbf{0.03}}& \res{0.92}{0.04}& \res{0.92}{0.02}& \res{0.92}{0.03}& \res{0.92}{0.04}& \res{0.92}{0.04}\\ 
 \textbf{KNN}&\textit{\textbf{0.05}}& \res{0.93}{0.02}& \res{0.94}{0.01}& \res{0.93}{0.02}& \res{0.93}{0.02}& \res{0.93}{0.02}\\ 
 &\textit{\textbf{0.07}}& \res{0.92}{0.00}& \res{0.92}{0.00}& \res{0.92}{0.00}& \res{0.92}{0.01}& \res{0.92}{0.00}\\ 
 &\textit{\textbf{0.09}}& \res{0.96}{0.01}& \res{0.95}{0.00}& \res{0.96}{0.01}& \res{0.96}{0.01}& \res{0.96}{0.01}\\ 
 &\textit{\textbf{Sup.}}& \res{0.97}{0.01}& \res{0.97}{0.00}& \res{0.97}{0.01}& \res{0.97}{0.01}& \res{0.97}{0.01}\\ \midrule

 &\textit{\textbf{0.01}}& \bestres{0.97}{0.01}& \bestres{0.97}{0.00}& \bestres{0.97}{0.01}& \bestres{0.97}{0.01}& \bestres{0.97}{0.01}\\ 
 &\textit{\textbf{0.03}}& \res{0.78}{0.09} &\res{0.78}{0.10} &\res{0.85}{0.05} &\res{0.78}{0.09} &\res{0.77}{0.10}\\ 
 \textbf{LR}&\textit{\textbf{0.05}}& \res{0.94}{0.02} &\res{0.94}{0.02} &\res{0.95}{0.02} &\res{0.94}{0.02} &\res{0.94}{0.02}\\
 &\textit{\textbf{0.07}}& \res{0.92}{0.07} &\res{0.92}{0.06} &\res{0.93}{0.05} &\res{0.92}{0.06} &\res{0.92}{0.07}\\
 &\textit{\textbf{0.09}}& \res{0.96}{0.01}& \res{0.96}{0.00} &\res{0.96}{0.01} &\res{0.96}{0.01} &\res{0.96}{0.01}\\ 
 &\textit{\textbf{Sup.}}& \res{0.96}{0.01}& \res{0.96}{0.00}& \res{0.96}{0.01} &\res{0.96}{0.01} &\res{0.96}{0.01}\\ \midrule

 &\textit{\textbf{0.01}}& \res{0.87}{0.03} &\res{0.87}{0.02} &\res{0.88}{0.02} &\res{0.87}{0.03} &\res{0.87}{0.03}\\
 &\textit{\textbf{0.03}}& \res{0.92}{0.02} &\res{0.93}{0.02} &\res{0.92}{0.02} &\res{0.92}{0.02} &\res{0.92}{0.02}\\ 
 \textbf{RF}&\textit{\textbf{0.05}}&\res{0.90}{0.02} &\res{0.90}{0.01} &\res{0.90}{0.02} &\res{0.90}{0.02} &\res{0.90}{0.02}\\
 &\textit{\textbf{0.07}}& \res{0.90}{0.03} &\res{0.91}{0.01} &\res{0.91}{0.02} &\res{0.90}{0.03} &\res{0.90}{0.03}\\
 &\textit{\textbf{0.09}}& \res{0.95}{0.01} &\res{0.96}{0.01} &\res{0.95}{0.01} &\res{0.95}{0.02} &\res{0.95}{0.02}\\ 
 &\textit{\textbf{Sup.}}& \res{0.97}{0.00} &\res{0.97}{0.00} &\res{0.97}{0.00} &\res{0.97}{0.00} &\res{0.97}{0.00}\\ \midrule

 &\textit{\textbf{0.01}}& \res{0.81}{0.05} &\res{0.82}{0.05} &\res{0.82}{0.05} &\res{0.81}{0.06} &\res{0.81}{0.06}\\ 
 &\textit{\textbf{0.03}}& \res{0.88}{0.04} &\res{0.88}{0.04} &\res{0.88}{0.04} &\res{0.88}{0.04} &\res{0.88}{0.04}\\ 
 \textbf{DT}&\textit{\textbf{0.05}}&\res{0.91}{0.01} &\res{0.92}{0.01} &\res{0.92}{0.01} &\res{0.91}{0.01} &\res{0.91}{0.01}\\ 
 &\textit{\textbf{0.07}}& \res{0.90}{0.02} &\res{0.91}{0.01} &\res{0.90}{0.02} &\res{0.90}{0.02} &\res{0.90}{0.02}\\ 
 &\textit{\textbf{0.09}}&\res{0.93}{0.01} &\res{0.93}{0.01} &\res{0.93}{0.01} &\res{0.93}{0.01} &\res{0.93}{0.01}\\ 
 &\textit{\textbf{Sup.}}& \res{0.95}{0.01} &\res{0.97}{0.00} &\res{0.95}{0.01} &\res{0.95}{0.01} &\res{0.95}{0.01}\\ \bottomrule
\end{tabular}
}
\label{tab_model}
\vspace{-3em}
\end{center}
\end{table}

The table shows that increasing labeled data does not necessarily improve the model's performance but increases its stability. Moreover, the results in the SSL mode do not differ significantly from supervised ones. This suggests that CT profiles share similar characteristics, as partially discussed in §\ref{sec:providers}, and algorithms can converge by taking a few labeled data. On the contrary, adding more samples could lead to over-fitting or biasing the classifier, reducing prediction accuracy (as happened for LR 0.03). The LR classifier with 1 percent of labeled data (penalty = \textit{l2}, C = 1, solver = \textit{liblinear}) showed the best cross-validation results among the semi-supervised models, so it was selected as the final model.\footnote{We discarded RF sup. (same scores) since the paper focuses on SSL. Practitioners should choose models with the best performance.} Such a model reached 95\% accuracy and F1-score on the test set and was used in the remainder of our analyses.


\subsection{Baseline Comparison}
\label{subsec.baseline}
To assess the quality of our results, we compared them with previous IG fake and bot account detection mechanisms~\cite{thejas2019learning, sheikhi2020efficient, akyon2019instagram, purba2020classification}. Only Akyon et al.~\cite{akyon2019instagram} released their data, so we used their dataset comprising authentic and fake/bot accounts to train all the baselines, adapting the features and re-implementing the models. Each baseline was tested on all our CT accounts (provider by provider) and real accounts. Table~\ref{tab:baselines} reports the avg$\pm$std in detecting CT profiles for each provider. Our algorithm outperforms all the baselines, being statistically better\footnote{Unpaired \textit{t} test with $\alpha = 0.05$ as significance threshold.} than the best baselines for Recall ($p$-value $<$ 0.05) and F1-score ($p$-value $<$ 0.01). The lower baselines' recall can be explained by CT accounts resembling real accounts characteristics, avoiding detection as expected. However, the relatively high standard deviations imply that the quality of CT providers varies significantly, i.e., some of them deliver lower-quality accounts, detectable by previous methods. The presence of low-quality profiles also highlighted in Table~\ref{tab:fake_providers}, allowed our detector to spot both CT and \textit{classic} fake accounts, making it more reliable than previous models trained on simple bots or synthetic data. 

\begin{table}[!ht]
\caption{Baseline comparison in detecting CT profiles.}
\centering
\begin{tabular}{cccc}
\toprule
\textbf{\textit{Baseline}} & \textbf{\textit{Precision}} & \textbf{\textit{Recall}} & \textbf{\textit{F1-Score}} \\
\midrule
Thejas et al.~\cite{thejas2019learning} & \res{0.77}{0.05} & \res{0.89}{0.09} & \res{0.82}{0.06} \\
Sheika et al.~\cite{sheikhi2020efficient} & \res{0.94}{0.02} & \res{0.84}{0.11} & \res{0.88}{0.07} \\
Akyon et al.~\cite{akyon2019instagram} & \res{0.87}{0.05} & \res{0.83}{0.19} & \res{0.84}{0.14}\\
Purba et al.~\cite{purba2020classification} & \res{0.92}{0.03} & \res{0.80}{0.14}  & \res{0.85}{0.09}  \\ \midrule
Our & \bestres{0.95}{0.02}& \bestres{0.95}{0.03} & \bestres{0.95}{0.02}
\\
\bottomrule
\end{tabular}
\label{tab:baselines}
\vspace{-1em}
\end{table}

\begin{figure}[h!]
    \centering
    \includegraphics[width=\columnwidth]{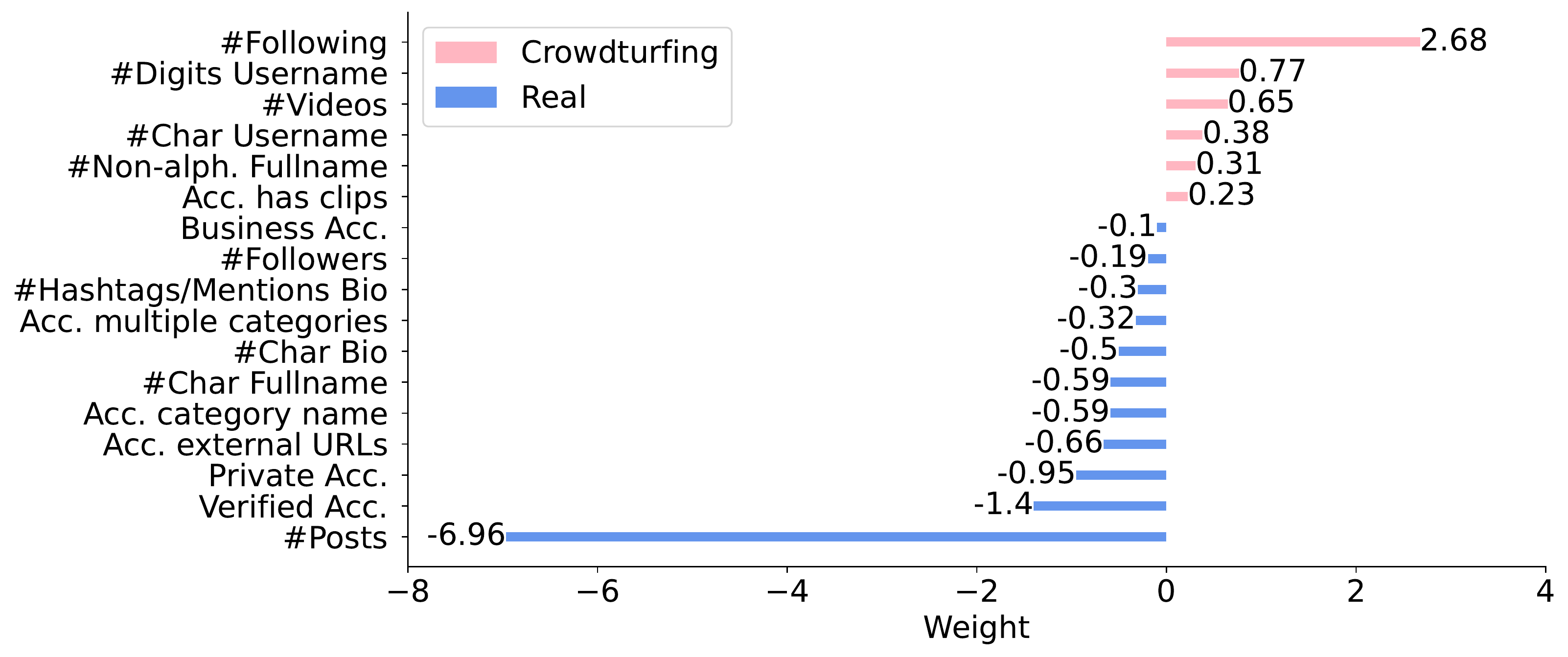}
    \vspace{-0.5cm}
    \caption{Logistic Regression weights to discriminate Crowdturfing (positive label) vs Real (negative label) profiles.} 
    
    \label{fig:feat_import}
    \vspace{-1.5em}
\end{figure}

We now explore the features' importance to explain why baselines performed worse. Figure~\ref{fig:feat_import} shows our model coefficients based on standardized features, so they are comparable. Baselines' most predictive features were the number of posts, following, followers, and bio length~\cite{thejas2019learning, sheikhi2020efficient, akyon2019instagram, purba2020classification}. While the number of following and posts is also crucial for us, the followers and bio length are less influential. The reason is that bots and simple fake accounts tend to have few followers and no bio, thus biasing baselines. Instead, CT profiles usually have many followers and genuine bios since they are real people profiles. Moreover, baselines do not leverage username and fullname characteristics, the number of videos, and if an account is private or verified, which are relevant to us.
This suggests our model performs better due to the training data that includes CT profiles and the features we extracted (e.g., \# digits in username) rather than the model itself. Nonetheless, we contribute to the state-of-the-art by demonstrating that (i) \textit{classic} fake accounts detectors are not enough to effectively detect CT profiles, (ii) the training data are more important than the detection algorithm, (iii) the task can be efficiently solved with SSL algorithms, significantly reducing (99\% less!) the time and costs to label data.

\section{Crowdturfing Analysis: Profiles Information}
\label{sec:profile_analysis}

With our CT profile detector trained, we are ready to analyze CT engagement in the wild. 
In our detection strategy, we detect profiles involved in CT using our model (§\ref{subsec:model_sel}) and label their engagement accordingly. 
Since CT profiles contribute to a fake engagement, we will also refer to them as \textit{fake} (non-genuine) accounts and engagement vs. \textit{real} (genuine) ones. 
For our analyses, we collected the comments and commenters' profile information\footnote{Profiles info were collected via Instaloader \url{https://instaloader.github.io/. }} of 50 recent posts of 20 mega-influencers with over 1 million followers (1000 posts in total). 
We selected posts at least five days old to allow IG automatically remove \textit{classic} fake interactions~\cite{ig_detectors, ig_authenth}.
The influencers come from different nationalities and the following categories: fashion, beauty, fitness, art, music, lifestyle, and family.
In total, we gathered 603,007 comments generated by 248,388 unique users. 
The reasons why we collected only comments-related information and e.g., not likes, are discussed in the comments analysis section (§\ref{sec:comment_analysis}).

Our CT detection model detected 55,719 CT profiles among the 248,388 collected ($\sim$22\%). This percentage aligns with the estimate of 20-40\% in celebrities' accounts~\cite{conti2022virtual}. 
We acknowledge that some of the detected accounts may not be CT; however, we are still dealing with ``advanced'' fake profiles that have bypassed (i) the automatic screening mechanisms employed by IG~\cite{ig_detectors, ig_authenth} (and Meta~\cite{apruzzese2023realgradients} in general), and (ii) the potential moderation done by the influencers themselves (e.g., by removing blatant spam comments).
Therefore, we can assume our further analyses will primarily focus on CT or advanced fake profiles that resemble and act as legitimate profiles. 
In this section, we provide a detailed study of CT profiles' information, including the number of followers and following  (§\ref{subsec:follower_followings}), biography 
(§\ref{subsec:fake_biography}), and external URLs (§\ref{subsec:fake_external_URL}), to determine whether CT profiles engage in malicious activities besides crowdturfing. 

\subsection{Followers and following ratio analysis}
\label{subsec:follower_followings}
\begin{figure}[h]
    \centering
    \includegraphics[scale=0.15]{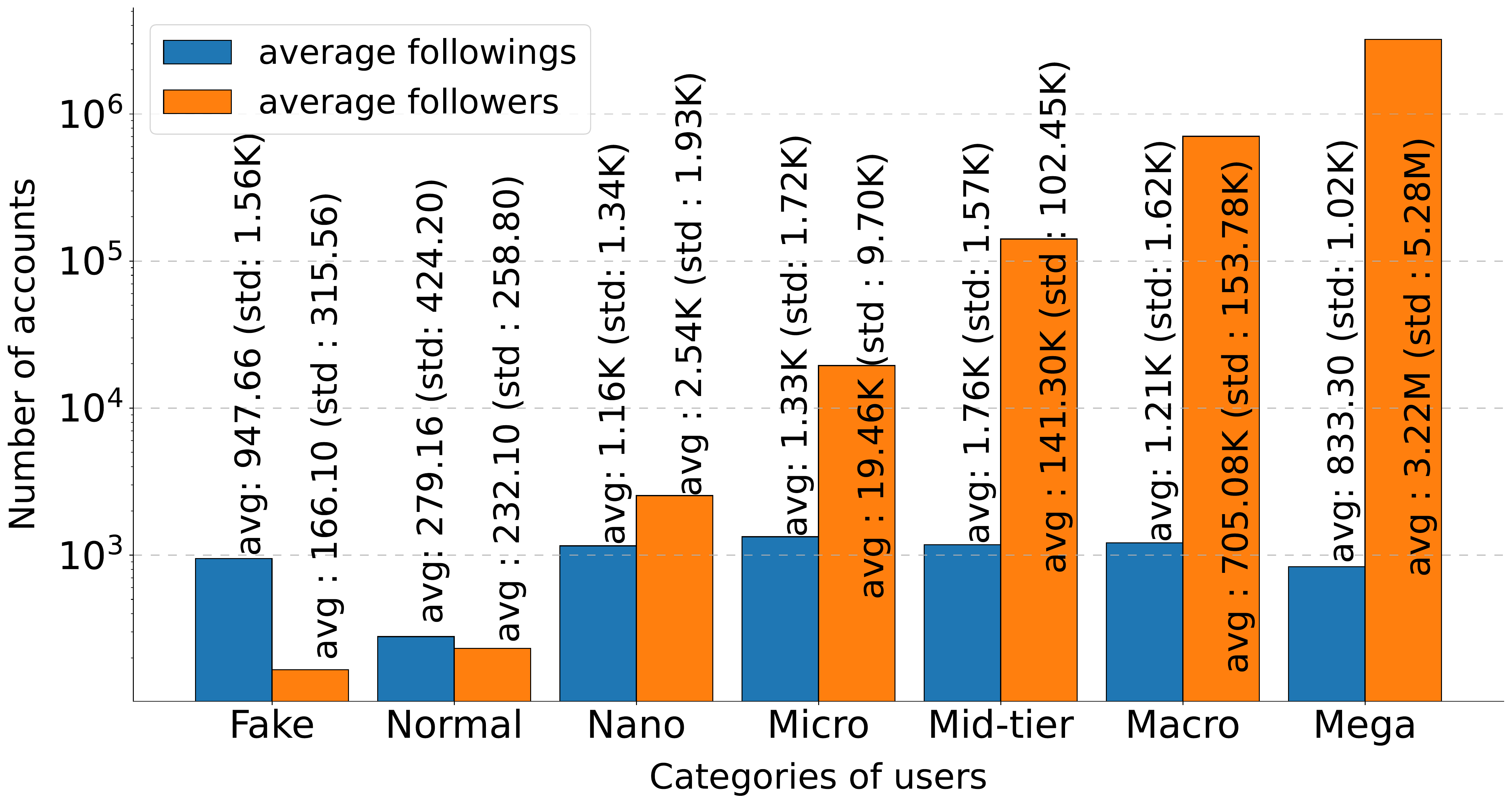}
    \caption{Followers and following avg and std of CT users (Fake) and different categories of real users. Y-axis is in log scale.}
    \label{fig:average_std}
\end{figure}

To increase other accounts' engagement (and therefore gain more money), a fake account will display an unusually high number of following (§\ref{sec:providers}). Conversely, genuine users should have a more balanced ratio of followers and following according to IG averages~\cite{ig_avgs}.
Figure \ref{fig:average_std} shows the mean and std of followers/following for fake and real users. Real users are divided into normal and influencers tiers\footnote{\url{https://www.shopify.com/id/blog/instagram-influencer-marketing}} as follows:  

\begin{itemize}
    \item {\itshape Normal accounts}: less than 1,000 followers;
    \item {\itshape Nano influencers}: $1,000\le \textrm{followers}<10,000$;
    \item {\itshape Micro influencers}: $10,000\le \textrm{followers}<50,000$;
    \item {\itshape Mid-tier influencers}: $50,000\le \textrm{followers}<500,000$;
    \item {\itshape Macro influencers}: $500,000\le \textrm{followers}<1,000,000$;
    \item {\itshape Mega influencers}: more than 1,000,000 followers.
\end{itemize}

\begin{figure}[h]
    \centering
    \includegraphics[scale=0.16]{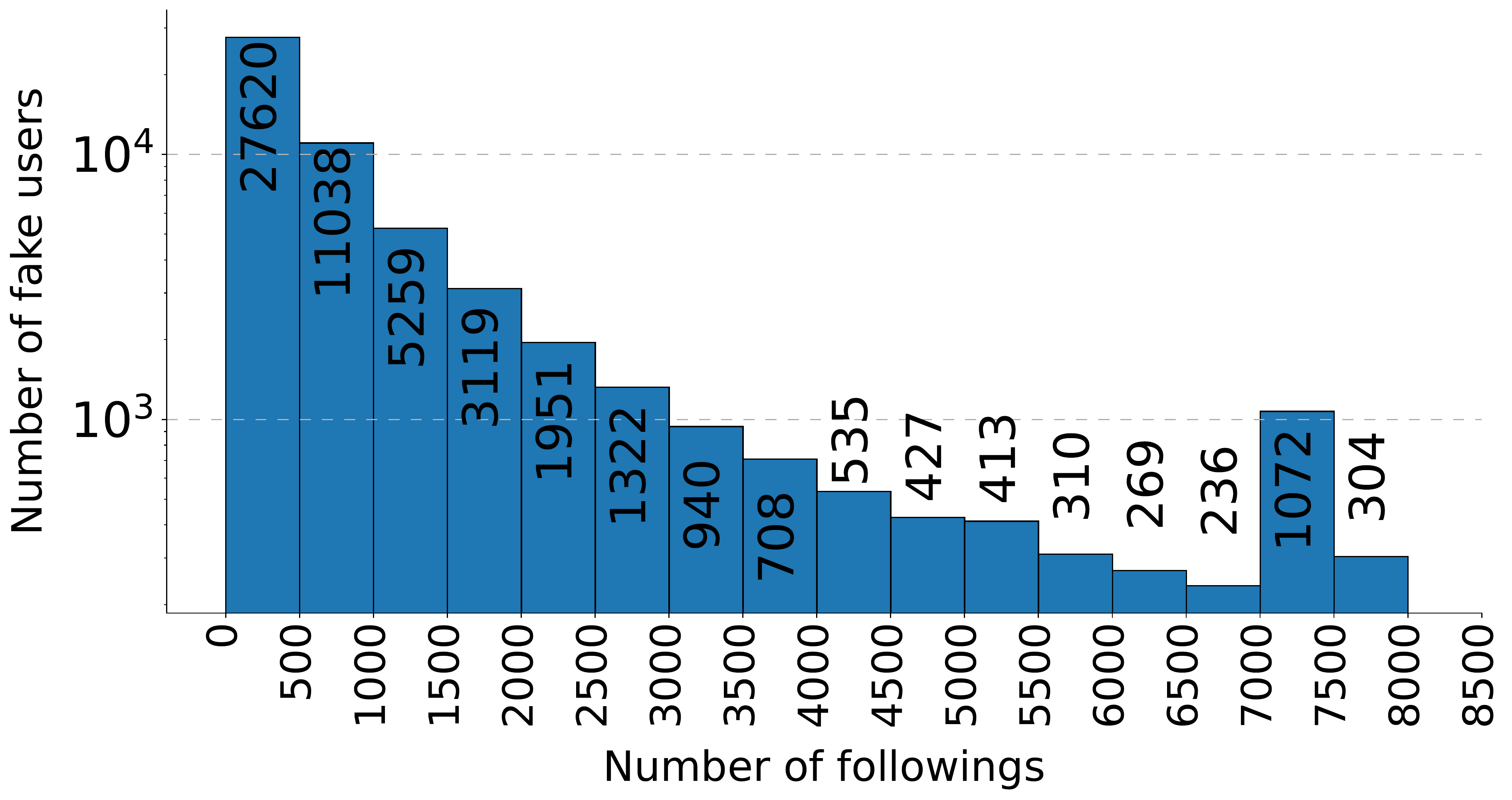}
    \caption{Distribution of fake accounts' following.}
    \label{fig:fake_following_distribution}
    \vspace{-0.5em}
\end{figure}

The graph shows that the number of followers of fake users is (on average) much smaller than the number of following. Indeed, these accounts are incentivized to follow more people to grow their earnings through CT activities, confirming our initial assumption. Following and followers of normal users are balanced, but as the popularity of the genuine account grows, followers increase exponentially while following hovers around 1000. For more popular influencers, the standard deviation increases simply because their categories include wider ranges (e.g., from one to hundreds of million followers for mega influencers). 
We further inspected the following distribution of CT accounts in Figure~\ref{fig:fake_following_distribution}.
Most CT accounts have between 0 and 500 following, with the number decreasing as the following increases, suggesting CT accounts tend to maintain a low profile to avoid being flagged as spammers. 
An exception occurs in the last two bins. IG introduced a 7500 following limit\footnote{\url{https://help.instagram.com/408167069251249}} to contrast spamming activities, and many CT (probably more similar to \textit{classic} fake) accounts are just below this limit.
Despite it, 304 fake profiles likely surpassed the threshold before its introduction.

\subsection{Fake profiles biography analysis}
\label{subsec:fake_biography}
Many \textit{classic} fake IG accounts use a catchy biography to lure victims into clicking malicious links. Thus, we tried to find suspicious words in the CT users' biographies. To this aim, we created a list of 31 elements, including words and emojis often used by this fake user. The list, based on our knowledge of fake behavior and a brief manual inspection, contained words like ``stories", ``chat", ``follow", ``gain", ``click", ``link", and emojis usually linked to malicious or sexual activities, like ``\includegraphics[width=0.3cm]{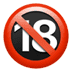}'', ``\includegraphics[width=0.3cm]{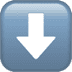}'',
``\includegraphics[width=0.3cm]{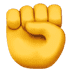}'',
``\includegraphics[width=0.3cm]{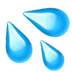}'',
``\includegraphics[width=0.3cm]{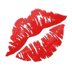}''~\cite{conti2022captcha}. Only 5635 CT accounts (10.11\% of the total detected) had at least one of the elements of the list. Thus, most CT accounts do not seek to boost their profiles or induce people to click links. Rather, they are interested in making profits by increasing the engagement of other accounts.

\subsection{Fake profiles external URLs analysis}
\label{subsec:fake_external_URL}
The last analysis performed on the CT accounts is based on their external URLs, aiming to understand the most used URLs among CT users and whether they could be vectors of attacks conducted over social networks~\cite{maliciousURLsocial}. Of the total fake accounts, only 2834 (5.08\%) had an external URL on their profile page. We grouped these 2834 URLs into the following categories: 

\begin{itemize}
    \item {\itshape Videogame}: Youtube, Twitch, Discord;
    \item {\itshape Messaging}: WhatsApp, Telegram;
    \item {\itshape Social Network}: Facebook, Twitter, Instagram, etc.; 
    \item {\itshape Music \& Photography}: Spotify, Soundcloud, Vsco.co;
    \item {\itshape Email \& Google services}: Gmail, Maps, Outlook;
    \item {\itshape URL redirecting}: Linktr.ee, Tinyurl, Linkr.bio, Bit.ly;
    \item {\itshape Shopping \& Payment}: PayPal, Vinted, Amazon, etc.;
    \item {\itshape Personal website \& Petition}: Blogspot, Wordpress, etc.;
    \item {\itshape Adult content}: URLs to different adult websites;
    \item {\itshape Other}.
\end{itemize}

\begin{figure}[ht!]
    \centering
    \hspace*{-0.7cm}
    \includegraphics[scale=0.20]{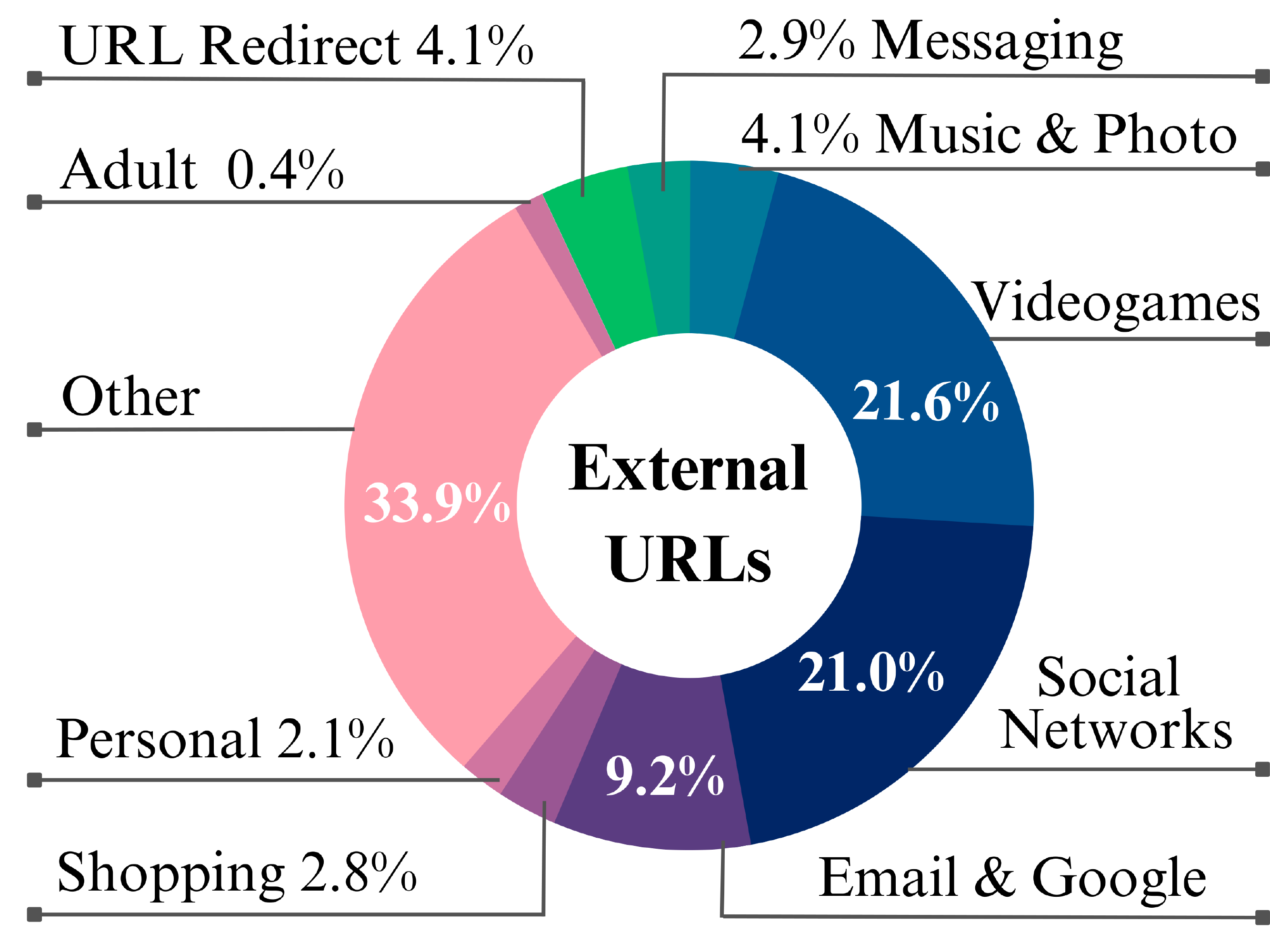}
    \caption{Categories of External URLs of the fake profiles.}
    \label{fig:fake_external_url}
    \vspace{-1em}
\end{figure}

Inside the categories, we also included shortened URLs (e.g., wa.me or t.me for WhatsApp and Telegram, respectively).
The results are shown in Figure~\ref{fig:fake_external_url}.
Remark that even if the categories contain well-known websites, some can be used for malicious purposes. For instance, we found many WhatsApp links starting a conversation or a phone call with strangers who could easily be scammers. Similarly, we inspected and monitored Telegram URLs, grouping them into:
\begin{itemize}
    \item {\itshape Conversation}: Similarly to WhatsApp URLs, starts a conversation with a potential scammer;
    \item {\itshape Piracy}: Illegal groups that share movies and tv series;
    \item {\itshape Selling}: Scam groups that try to sell clothes, Amazon gift cards, cryptocurrencies, NFTs, etc. 
\end{itemize}

Moreover, \textit{classic} fake profiles commonly use redirect URLs to route the victim to a malicious site~\cite{linktreemalicious}.
From Figure~\ref{fig:fake_external_url}, it is possible to see that the ``Other" section is more relevant than the other categories inside the pie chart, with precisely 961 URLs. It contains very heterogeneous URLs, making their categorization challenging. To better understand these URLs' nature (i.e., if they are malicious), we have relied on a fraud prevention and detection service called Ipqualityscore.\footnote{\url{https://www.ipqualityscore.com/}} It allows checking for suspicious links by using a mixture of blacklists and deep learning algorithms, and to define the following URLs categories:

\begin{itemize}
    \item {\itshape Parked}: Domains that have been dormant for a long time; 
    \item {\itshape Spamming}: Websites that spams malicious content;
    \item {\itshape Malware}: Websites hosting viruses, malware, etc.;
    \item {\itshape Phishing}: Websites hosting fake login, or sign up forms;
    \item {\itshape Adult}: Websites that contain adult content.
\end{itemize}

\begin{figure}[ht!]
    \centering
    \hspace*{-0.7cm}
    \includegraphics[scale=0.15]{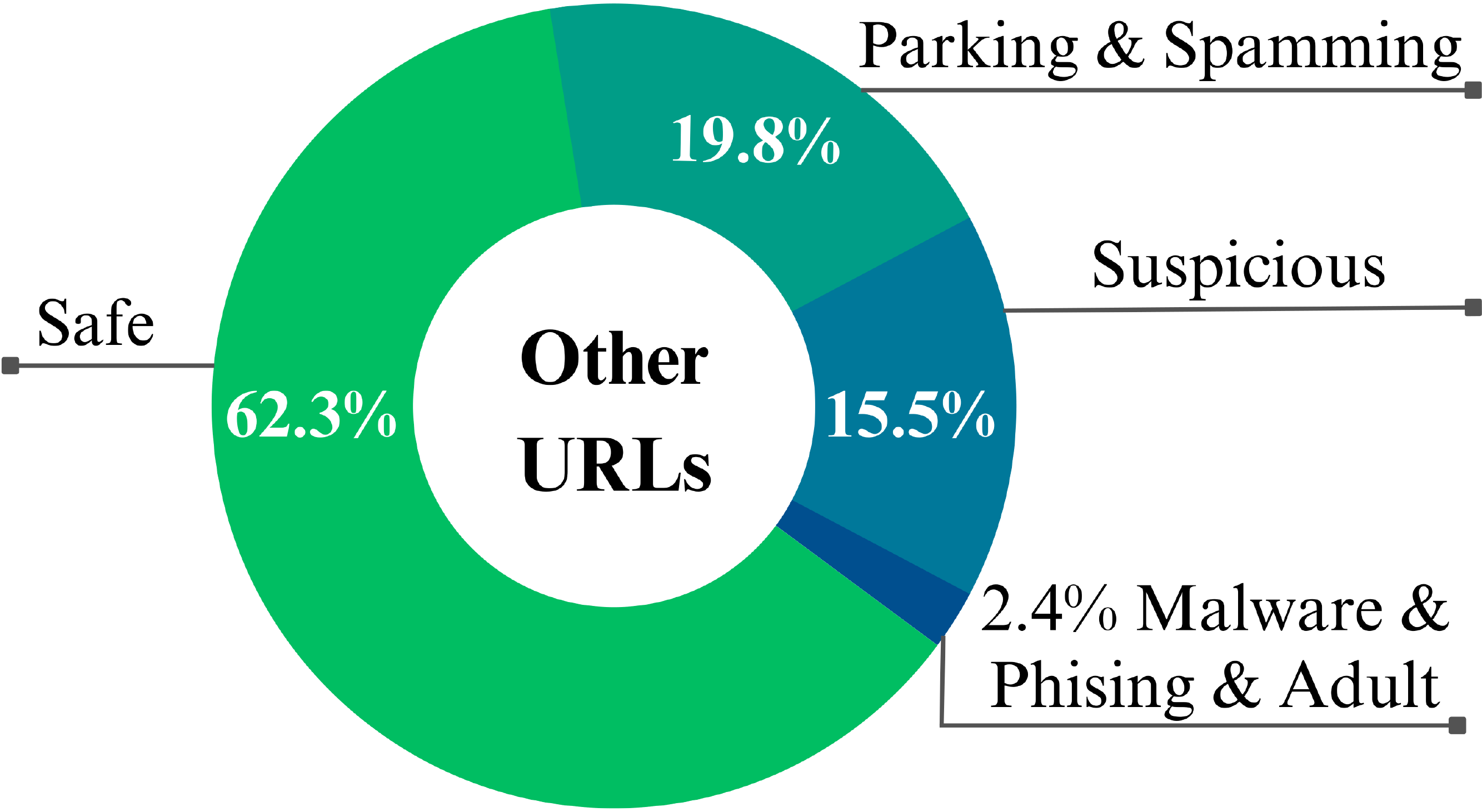}
.    \caption{Results provided by the fraud prevention and detection service on the URLs in the ``Other" category.}
    \label{fig:other_url_study}
\end{figure}

The results of this evaluation are shown in Figure~\ref{fig:other_url_study}.
For convenience, we grouped the ``Phishing", ``Malware", and ``Adult" categories since they had very few matches. From the total 961 ``Other'' URLs, 599 were considered safe, while the remaining 362 were divided as follows:
\begin{itemize}
    \item 190 URLs were parked and/or spamming websites;
    \item 5 URLs were marked as malware websites;
    \item 7 URLs were marked as phishing websites;
    \item 11 URLs were adult websites;
    \item 149 were considered suspicious websites.
\end{itemize}
These results show that most external URLs in the ``Other" category were considered safe. However, many spamming and suspicious websites can be used for malicious purposes. Comparing the obtained results to the overall number of CT users, we can confirm that most are solely involved in CT activities rather than malicious activities.

\section{Real vs Crowdturfing Comments Analysis}
\label{sec:comment_analysis}
This section analyses CT engagement. In particular, we aim to understand if CT can be directly spotted by actions (e.g., comments) instead of leveraging profile information. As stated before, CT profiles are driven by humans, so intuitively, there should be little to no difference between real and fake engagement, but we cannot draw conclusions without proper analysis.
On IG, the primary forms of engagement are liking and commenting. CT likes cannot be isolated from the action itself since it carries no information beyond temporal data (unavailable on IG). Instead, comments provide valuable information (e.g., stylometric features) that could be used for CT detection. Moreover, comments present a higher level of public expression than likes~\cite{aldous2019view} and are considered more important to boost the visibility of an account~\cite{instaalg2022,instaalg}. For these reasons, we focus on comments in this section, presenting five studies to spot the differences between comments made by CT and real users. 

\subsection{Stylometric Analysis}
\label{sub:stylometric}
From our dataset, we isolated 121,822 comments shared by CT users and 481,185 from legit ones. We performed a stylometric analysis similar to the one conducted in~\cite{bhargava2013stylometric}, based on Lexical Features, Syntactical Features, and Emoji Features.

\textbf{Lexical Features.}
\label{subsub:lexical}
We calculated the number of sentences per comment, the number of words in each comment, the number of words in each sentence, and the length of the comments. We found several statistically significant ($p$-value $<$ 0.001) differences: CT users have an overall mean of 1.13 words per comment, while the real ones have 4.34. Similarly, the number of words per sentence is 0.94 for the CT accounts and 2.96 for the real ones. Instead, both categories of users have a mean of 1.35 sentences per comment. In each comment, there is a low repetition of words: we obtained that 99\% of them, made by CT users, have no word repetitions, while for the legit users is 97\%. 
Another important distinction is the length of the comments: the CT users shared text with a mean length of 28.89 (std: 61.19) characters (emojis included), while the legit users have a mean of 23.74 (std: 46.74). Even if similar, they are statistically significant ($p$-value $<$ 0.001). The emoji comparison better explains how real users, with more words, have shorter comments.

\textbf{Syntactical Features.}
We counted the number of comments starting with a capital letter, punctuation present in the text, and capital words. We found very close results between CT and real users: 
the beginning of the comment is in uppercase for 33.86\% of comments made by CT users and for 34.94\% of real ones.
35\% of the comments have some punctuation for both accounts categories. 
Finally, we saw that both categories do not use upper-cased words: the mean of the ratios between uppercase words and all the words in each comment are 0.021 for the CT users and 0.025 for the legit ones.

\textbf{Emoji Features.}
\label{subsub:emoji}
We detected emojis in the comments using demoji\footnote{\url{https://pypi.org/project/demoji/}}. 
Our study focused on the presence of emojis and alphanumerical text in the comments, in particular:
\begin{enumerate}
    \item \textbf{The percentage of comments with at least one emoji};
    \item \textbf{Most used emojis}: the percentage of an emoji among all the fake comments. Multiple occurrences of the same emoji on the same comment increase the counter by one.
    \item \textbf{Avg emojis when present}: considering only comments presenting emojis, the avg number of them. Multiple occurrences of the same emoji increase the counter accordingly.
\end{enumerate}


Table~\ref{tab:style_res} reports the results. The top-most emojis used are equal for both users, with similar percentages. Another meaningful result is that even if real users have, on average, slightly more comments with emojis
, the quantity of emojis in such comments is fewer compared to CT users. 
This result might explain the outcomes on comment length found in §\ref{subsub:lexical}. To sum up, results obtained so far show some stylometric differences, but mostly similarities between CT and real users when the focus is on emoji used, sentences per comments, or syntactical features. Legit users share comments with more words, fewer emojis, and an overall shorter comment length.

\begin{table}[!ht]
    \centering
    \caption{Emojy-based Stylometric analysis. CE = Comments with Emoji, EPC = Avg Emoji per comment.}
    \resizebox{0.8\columnwidth}{!}{
    \begin{tabular}{cc|cccccc|c}
    \toprule
        & \textbf{\textit{CE}} & \multicolumn{6}{c|}{\textbf{\textit{\% of Most used Emoji}}} & \textbf{\textit{EPC}} \\
        & (\%) & \includegraphics[width = 0.4cm]{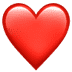} & \includegraphics[width = 0.4cm]{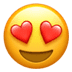} & \includegraphics[width = 0.4cm]{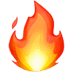}  & \includegraphics[width = 0.4cm]{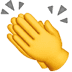} & \includegraphics[width = 0.4cm]{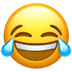} & \includegraphics[width = 0.4cm]{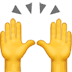}& \\
        \midrule
         Fake & 71.6  & 25.18 & 19.92 & 10.57 & 4.91 & 4.03 & 2.73 & 3.557 \\
         Real & 72.7 & 22.30 & 18.46 & 14.42 & 5.00 &  4.92 & 3.04 & 3.211 \\
         \bottomrule
    \end{tabular}
    }
     \label{tab:style_res}
     \vspace{-1.5em}
\end{table}

\subsection{Common Words Analysis}
\label{sub:cwa}
\begin{figure}[!h]
    \centering
    \subfloat[\centering Fake Users]{
        \includegraphics[width=0.35\linewidth]{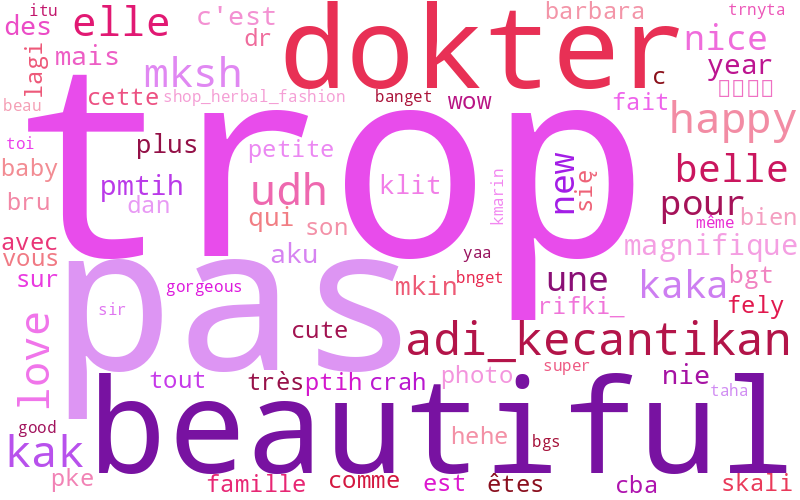}
        \label{fig:wc_fake_comments} }
    \hfil
    \subfloat[\centering Real Users]{
        \includegraphics[width=0.35\linewidth]{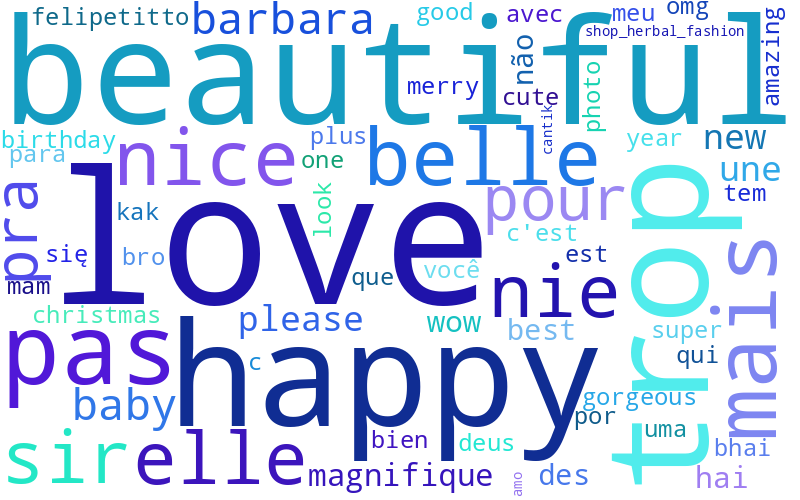}
        \label{fig:wc_real_comments} }%
    \caption{Most used words by fake and real users.}
    \label{fig:wordclouds}%
    \vspace{-1em}
\end{figure}

We analyzed the most common words CT and real profiles use.
As a pre-processing, we removed emojis, punctuations, and unproductive words with less than three characters, e.g., ``and", ``the", ``you".
The word clouds in Figure~\ref{fig:wordclouds} show fake and real users' top 100 most used words.
In general, we found a lot of positive and loving expressions, such as beautiful, love, happiness, niceness, etc.

%

An interesting word from Figure~\ref{fig:wc_fake_comments} is ``Dokter", which appeared in 1069 comments.
By investigating the accounts spamming this word, we might have found a botnet whose objective is to spam ``IG doctors'' accounts. All these doctors' profiles have a WhatsApp business link starting a chat with a message to complete: 
``*NAME*: *CITY/STATE*: *ORDER/COMPLAINTS*: *AGE*:". 
Some doctors' accounts no longer exist on IG, suggesting they probably violated the ToS.
Other similar accounts had the format ``\verb|dr.[doctor_name]|",
presenting the same WhatsApp link and conversation, but different phone numbers. 
We found 1370 comments coming from 33 different accounts containing such words, suggesting the presence of a bigger malicious network.

\subsection{Number of Comments per User}
\label{sub:num_of_comments}
248,388 unique users posted the 603,007 comments we analyzed; thus, many users posted multiple comments. 
We found that a legit user, on average, has posted 1.95 comments (std 5.94), while a CT user has posted slightly more (2.24, std 7.57).
The result obtained in this analysis complies with the one in §\ref{sub:stylometric}: a CT user has a similar behavior as the legit user. However, a CT user generally shares more comments than a real one because their purpose is to generate engagement.
But to avoid IG bot detection, an account has to act like a real human being. 

\subsection{Language Analysis}
\label{sub:lang_analysis}
\begin{figure}[!h]
    \centering
    \subfloat{%
        \includegraphics[width=0.47\linewidth]{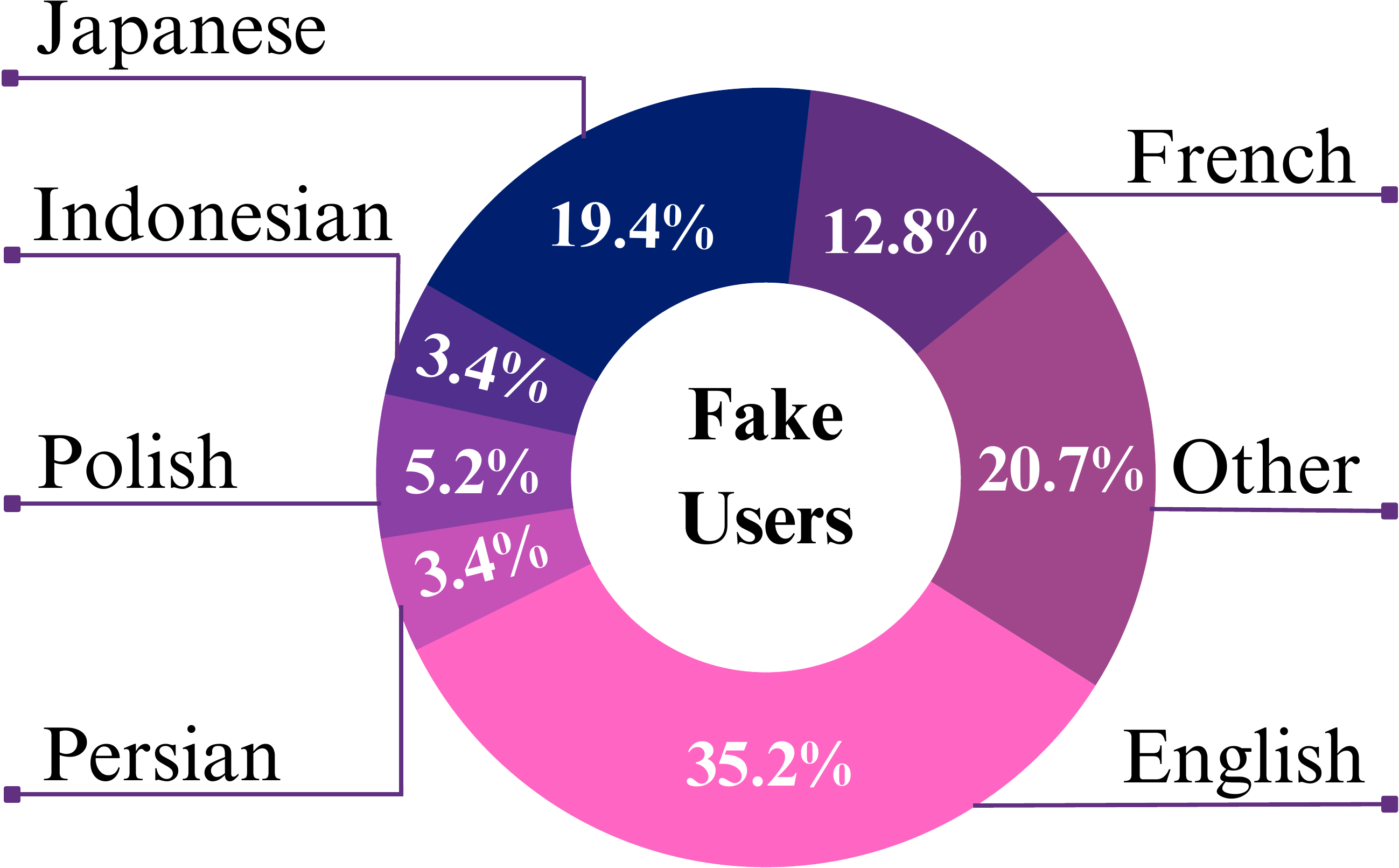}
        \label{fig:pie_lang_fake} }
    \subfloat{%
        \includegraphics[width=0.47\linewidth]{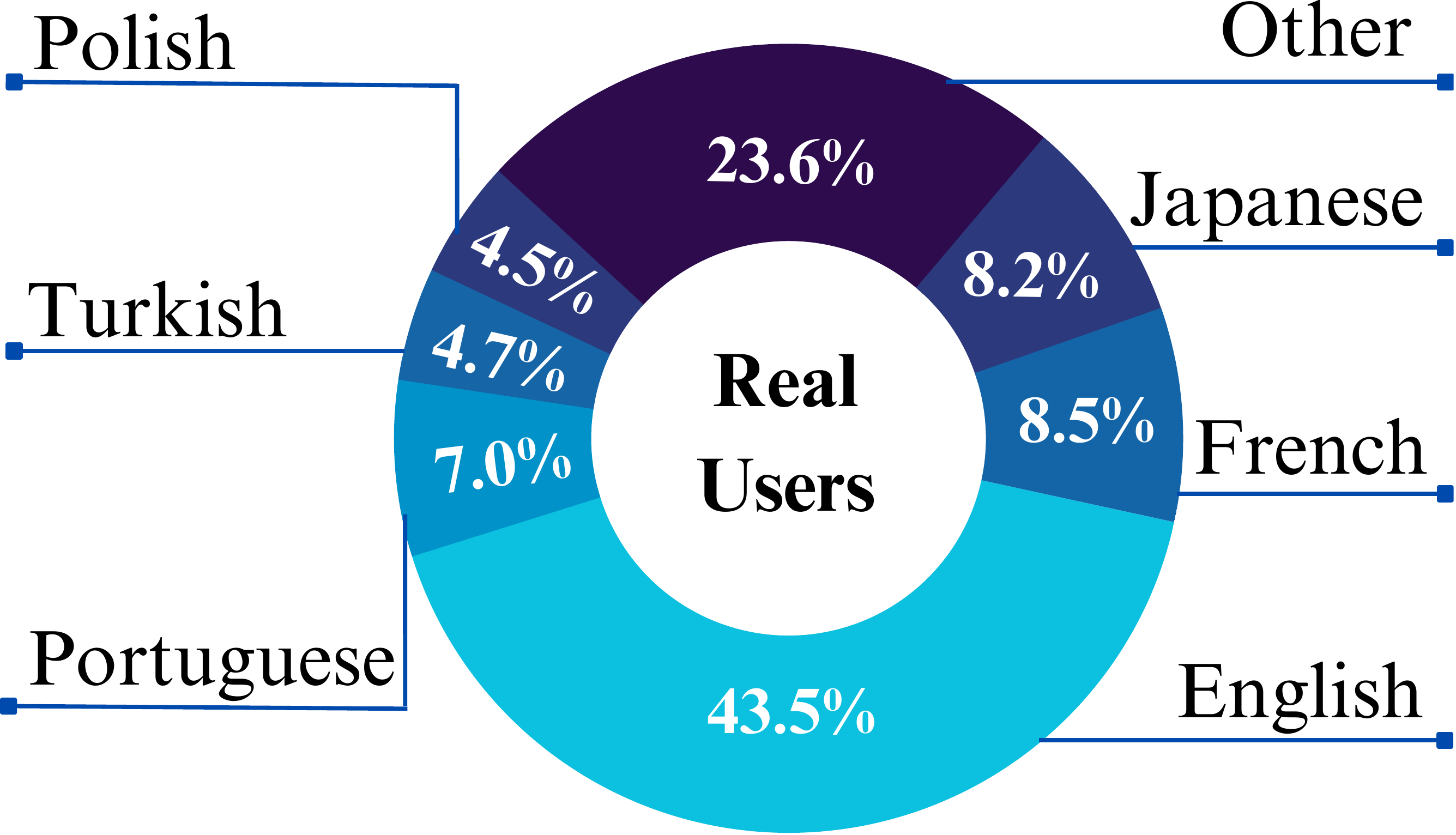}
        \label{fig:pie_lang_real} }%
    \caption{Languages detected in comments.}%
    \label{fig:languages}%
    \vspace{-1em}
\end{figure}

We analyzed the language used by CT and real users using SpaCy.\footnote{\url{https://spacy.io/usage/facts-figures}} 
The text was filtered out of emojis and then used as input for the neural network. The results are shown in Figure~\ref{fig:languages}.
In both CT and real comments, we found that the prominent language is English (35.2\% and 43.5\%, respectively), followed by Japanese and French.
The ``Other'' slices include more than 100 languages, each with a presence below 2\%.
They are probably the second largest sections of the pie charts because many comments just mentioned other accounts or used single words, complicating the language detection process.
Besides that, CT users likely adopt the language of their target community or, more commonly, English. In fact, as stated in §\ref{sec:providers}, many CT providers allow the option to deliver followers from specific geographical locations.

\begin{table*}[!ht]
\caption{Top-10 topics extracted from fake and real comments.}    
\scriptsize
\begin{tabular}{cp{5cm}p{1.8cm}|cp{5cm}p{1.8cm}}
\toprule
\multicolumn{3}{c|}{\textbf{Fake Comments}} & \multicolumn{3}{c}{\textbf{Real Comments}} \\
\textbf{\textit{N. Comm.}} & \textbf{\textit{Top Words}} & \textbf{\textit{Label}} & \textbf{\textit{N. Comm}} & \textbf{\textit{Top Words}} & \textbf{\textit{Label}} \\
\midrule

3817 & beautiful gorgeous sexy perfect hot amaze girl & Female Beauty & 13034 & beautiful gorgeous nice cute pretty lovely girl & Female Beauty \\
2290 & love beautiful cute smile god woman world girl & Love (woman) & 7547 & love good smile congrats great brother bro wish & Love (Males) \\
2117 & good want video well thank man bro life bike work & Man Compliment & 6983 & dream make want come time good life day hope & Life Dreams \\
1755 & happy birthday halloween republic thanksgiving & Pagan holidays & 6274 & christmas merry god bless family thank bible & Christmas \\
1476 & please christmas merry story follow check thank & Christmas/Follow & 6035 & happy new year birthday day family love republic & Pagan holidays \\
1136 & help fire turkey people stop please give helpturkey & Help Turkey & 6008 & help need fire people turkey please animal world & Turkey/Ecologists \\
1021 & trop wanna kiss lip red face belle pretty liplock & Kiss \& Face & 5658 & picture crazy bro think top video sick bike man & Exalt Men \\
674 & arm chest belly waist neck armpit thigh dance & Body parts & 5524 & follow check story post page like support profile & Follow/support \\
514 & problem solution wife money call whatsapp expert & Problems/Ads & 4846 & please congrats reply check story dance song real & Music \\
223 & love back help massage oil bubbs real magic  & Relax & 1381 & problem belle family life help solution marriage & Family Problems\\\bottomrule
\end{tabular}
\label{tab:topicsinference}
\vspace{-1.5em}
\end{table*}

\subsection{Topics Analysis}
\label{sub:topic_analysis}
To further investigate the behavior of CT and real users, we inspected the topics in their comments. Many state-of-the-art topic modeling algorithms, such as Latent Dirichlet Allocation (LDA), require long text to extract topics. However, social network comments are usually concise sentences, making the topic modeling more challenging. In our experiments, we used GPU-PDMM\cite{gpu-pdmm}, which is typically adopted to extract topics of tweets. Based on the Poisson-based Dirichlet Multinomial Mixture (PDMM) model, GPU-PDMM promotes the semantically related words under the same topic during the sampling process by using the Generalized Polya Urn (GPU) model. We considered only English comments for the analysis, after removing non-alphabetical characters, emojis, stop words, words shorter than three characters, and applying lemmatization. From our comments, 15,023 CT comments and 63,290 Real comments were suitable for the study. 
We instructed the model to distinguish ten topics in an unsupervised fashion, returning for each comment the belonging topic and the top words associated with each topic.
The results of the topics inference are shown in Table~\ref{tab:topicsinference}.

As expected, we find high alignment between topics covered by CT and real profiles. Most comments exalt female beauty, using compliments, love words, or positive feelings to boost engagement. In particular, CT comments contain more exaggerated terms, such as ``sexy'', ``perfect'', or ``amaze''. Conversely, we found few advertisement comments, likely to avoid being flagged as spammers. An interesting difference between CT and real comments is how they dealt with the \textit{Help Turkey} topic. For real profiles, we found additional words such as ``animal'' and ``world'', suggesting they also brought up other environmental arguments, while CT did not. For real comments, we also found a \textit{Follow/support} topic, which could be a false positive (some spammers were not detected) or that they did not care about being labeled as spammers. In summary, the topic analysis revealed some differences, but not consistently enough to allow for proper differentiation.

\section{Conclusion}
\label{sec:conclusion}
In this work, we developed an algorithm that leverages profiles' characteristics through semi-supervised learning to spot 
IG crowdturfing activities. To train our classifier, we purchased CT profiles from 11 providers, which we further studied to understand their services and the type of profiles involved in them. Our Logistic Regression classifier scored 0.95\% F1-score. 
To spot IG CT activities in the wild, we targeted the most recent posts of 20 influencers of different nationalities and categories. We mainly focused on comments, as they are a crucial engagement metric for accounts' visibility, and carry more information than likes. For this purpose, we collected 603,007 comments among the different posts made by 248,388 unique users. 
Our model labeled 55,719 of these profiles as CT accounts. We compared CT profiles and comments with genuine ones, concluding that CT activities would be difficult to detect based only on their activities. Indeed, CT profiles are mostly real profiles guided by real humans; thus, their activities are close to genuine ones. In contrast to bots or fake profiles, they seem to not be involved with malicious activities besides boosting other accounts' engagement.
In the future, we plan to distinguish between CT profiles and other ``advanced'' fake profiles we might have (in)voluntarily encountered in our analyses.
While IG and the research community focused a lot on detecting bots and automated accounts, we believe more studies should be conducted on CT activities or in general, advanced fake profiles which negatively impact influencer marketing, IG, and most of its users.



\section*{Ethical Considerations}
\label{subsec:ethical}
We faced two main ethical challenges: CT activities' involvement and data collection on IG. Our experiments were designed
following the exemption guideline from a formal review by our institute's IRB.
To deal with CT activities, we acted similarly to previous works that analyze underground activities~\cite{thomas2013trafficking, song2015crowdtarget, voronin2018crowdturfing}, first by dealing only with a small number of CT followers and platforms, minimizing our effect on them and IG. Second, we linked the followers to freshly created accounts that had no prior connection with other IG accounts, and we deleted them at the end of the study. Thus, CT activities were not involving legitimate users.

For data collection, we gathered only profiles' information and comments publicly available, removing all the information linked to individual subjects (e.g., name, profile picture). Similar to previous works~\cite{quercia2011mood, hanson2013tweaking}, we could not request informed consent to prevent participants from (in)voluntarily changing their behavior, causing the Hawthorne effect~\cite{franke1978hawthorne}. Since IG APIs do not return all the public information of a user's profile, yet visible by simply browsing it, we collected such data in an automated way, which is not allowed by the ToS. However, as argued by Fiesler et al.~\cite{fiesler2020no}, 
``ethical decisions regarding data collection should go beyond ToS and consider contextual factors of the source and research''. In particular, IG ToS allows manual collection, suggesting that automated collection is probably not allowed to avoid heavy servers' workload~\cite{fiesler2020no}. Therefore, we tuned our tools to collect data at a slow human-like pace, using only our 11 profiles over five months, avoiding any ban from the platform.


\balance 
\bibliographystyle{abbrv}
\bibliography{references}

\end{document}